# Design and test results of different aluminum coating layers on the sCMOS sensors for soft X-ray detection

W.X. Wang,[a] Z.X. Ling,[a,b,1] C. Zhang,[a,b] W.M. Yuan,[a,b] and S.N. Zhang[a,b,c]

[a] *Key Laboratory of Space Astronomy and Technology, National Astronomical Observatories, Chinese Academy of Sciences,*
  *Beijing, 100101, China*

[b] *School of Astronomy and Space Science, University of Chinese Academy of Sciences,*
  *Beijing, 100049, China*

[c] *Institute of High Energy Physics, Chinese Academy of Sciences,*
  *Beijing, 100049, China*
  *E-mail*: `lingzhixing@nao.cas.cn`

ABSTRACT: In recent years, tremendous progress has been made on complementary metal-oxide-semiconductor (CMOS) sensors for applications as X-ray detectors. To shield the visible light in X-ray detection, a blocking filter of aluminum is commonly employed. We designed three types of aluminum coating layers, which are deposited directly on the surface of back-illuminated sCMOS sensors during fabrication. A commercial 2k × 2k sCMOS sensor is used to realize these designs. In this work, we report their performance by comparison with that of an uncoated sCMOS sensor. The optical transmissions at 660 nm and 850 nm are measured, and the results show that the optical transmission reaches a level of about $10^{-9}$ for the 200 nm aluminum layer and about $10^{-4}$ for the 100 nm aluminum layer. Light leakage is found around the four sides of the sensor. The readout noise, fixed-pattern noise and energy resolution of these Al-coated sCMOS sensors do not show significant changes. The dark currents of these Al-coated sCMOS sensors show a noticeable increase compared with that of the uncoated sCMOS sensor at room temperatures, while no significant difference is found when the sCMOS sensors are cooled down to about -15°C. The aluminum coatings show no visible crack after the thermal cycle and aging tests. Based on these results, an aluminum coating of a larger area on larger sCMOS sensors is proposed for future work.

KEYWORDS: sCMOS; Back-side illumination; X-rays; optical blocking filter.

---

[1] Corresponding author.

# Contents

## 1. Introduction

Silicon semiconductor sensors are commonly used in soft X-ray detection, where optical blocking filters are employed to shield the influence of visible light. Free-standing aluminum-coated polymer films have been used in a number of X-ray satellites, such as Chandra [1], XMM-Newton [2], Swift [3] and Suzaku [4]. These films are typically less than 1 μm in order to transmit soft X-rays and thus mechanically fragile. Aluminum has also been directly deposited on the surface of the detector as an optical blocking filter in MAXI-SSC [5], Arcus [6] and REXIS [7], which is more mechanically robust than free-standing aluminum-coated polymer films. Since the pioneering application on the ASCA mission, CCD detectors have become the workhorse for X-ray telescopes. During the last several years, tremendous progress has been made on scientific CMOS (sCMOS) imaging sensors, which have a similar working principle to CCD sensors. Some new proposed X-ray missions, such as Einstein Probe [8] and Theseus [9], [10], [11], will use sCMOS detectors as their focal plane instruments.

We designed three different aluminum film coatings as the optical blocking filters: a full layer with a thickness of 100 nm, a full layer with a thickness of 200 nm, and a slits layer with a thickness of 200 nm. For the initial verification test of the performance of these coatings, we used a 2 cm × 2 cm back-illuminated sCMOS sensor to verify their performance.

In this paper, the design of the aluminum coatings and the sCMOS samples are described in section 2, the optical transmission results and their basic performance are reported in section 3 and section 4, and the summarization is presented in section 5.

## 2. Design of the coating layers and the sCMOS samples

To reduce the influence of visible light during X-ray detection, an optical blocking filter, usually composed of aluminum, is required. The thickness of an optical blocking filter depends on the environment used and the observation target. For example, 40 nm and 80 nm aluminum layers are used as the thin and medium filters on the XMM-Newton Observatory [2]. To achieve a level of optical transmission of less than $10^{-6}$, an aluminum film about 200 nm in thickness is required to be coated on the surface of the back-illuminated sCMOS in this work. Since the sCMOS sensors are primarily used in low temperature environments, and the coefficient of thermal expansion (CTE) is different between the aluminum coating and the material of the sCMOS sensor, two different patterns of aluminum coatings are designed to be sputtered on the surface of the sCMOS sensors. As shown in Figure 1, one is a full aluminum coating on the sensor. The other is designed with some slits with a width of 15 μm on the aluminum film, and these slits divide the aluminum film into different regions. These slits are intended to release the stress caused by the material with different CTE. Meanwhile, a 100 nm aluminum layer is also designed to be deposited on a sCMOS sensor for comparison of the optical attenuation performance. To avoid light penetrating into the photosensitive regions from the edges of the silicon die, which has been found to be the case in CCD [5], [12], [13], the covered area of the aluminum film is extended by 70 μm on each side, compared with the sensitive area of the sCMOS as shown in Figure 2.

The sCMOS sensors used in this work are the standard GSENSE400BSI, which is a mature product. A series of GSENSE400BSI products have been tested by several groups, and the results show that they are applicable for X-ray imaging spectroscopy [14], [15], [16], [17]. GSENSE400BSI is a back-illuminated sCMOS sensor, and has an active imaging pixel array of 2048 × 2048 with a pixel size of 11 μm × 11 μm. The epitaxial layer depth of the sensor used in this work is 10 μm. The maximum frame rate is 24 fps in HDR mode and 48 fps in STD mode. The readout noise is 2 e-. The photon-response non-uniformity is less than 2%. The pictures of these three kinds of Al-coated GSENSE400BSI sensors are shown in Figure 3.

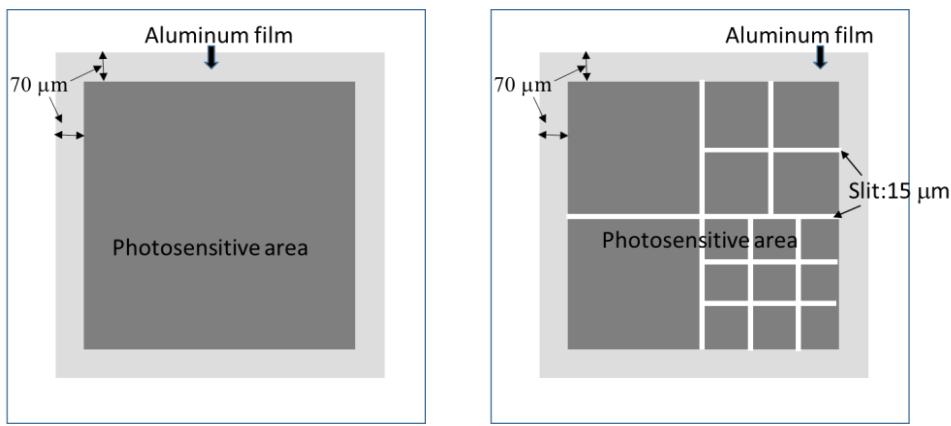

**Figure 1.** Schematic diagrams of the aluminum coating on the surfaces of the sCMOS sensors. (left) pattern 1: fully-aluminized; (right) pattern 2: aluminum coating with 15 μm slits which divide the imaging area into 15 regions.

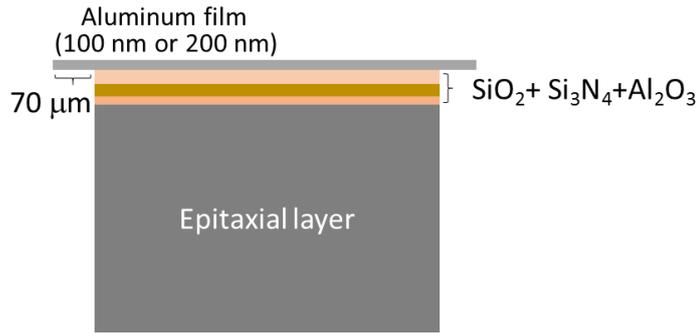

**Figure 2.** Schematic diagram of the cross section of the Al-coated sCMOS sensor.

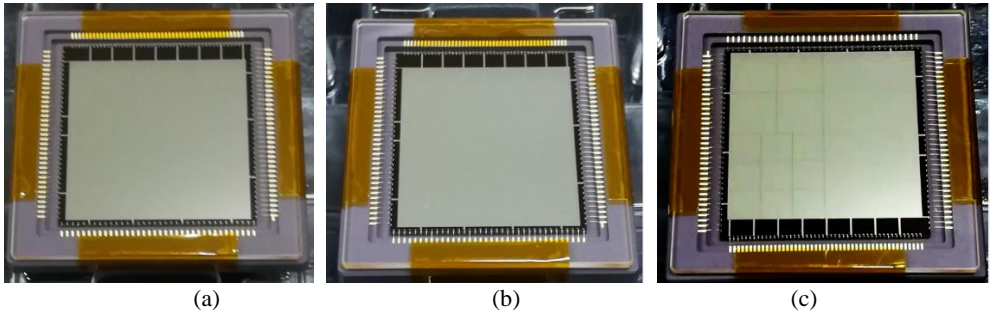

**Figure 3.** (a) sCMOS sensor fully coated with 100 nm Al; (b) sCMOS sensor fully coated with 200 nm Al; (c) 200 nm Al-coated sCMOS sensor with 15 μm slits.

## 3. Optical transmission measurements

We tested the optical transmission of these Al-coated sCMOS sensors at wavelengths of 660 nm and 850 nm. Two LED panel sources of 100 W each are used in the experiment. These LED sources are placed 40 cm away from the CMOS sensors, and the light intensity is about a few dozen watts per square centimeter monitored by a photodiode power sensor. Figure 4 shows the image taken by the 200 nm fully Al-coated sCMOS sensor with a 660 nm light source. The four sides of the sensor show an obvious light leakage. This phenomenon is similar to that observed on CCD sensors. This result proves that the 70 μm extension is not enough to block the light entering from the sides. Therefore, the central 1000 × 1000 pixels region, marked as the white line area in Figure 4, is used to test the optical transmission to avoid the edge effect. The non-uniformity of the light reaching this central area is less than ±3%, evaluated with the uncoated sCMOS sensor. Under the same light source, the signal of each pixel of the Al-coated sCMOS sensor divided by the mean signal of the uncoated sCMOS sensor gives the optical transmission. The histogram of the optical transmission of these pixels is shown in Figure 5, where the red lines show the results tested with 660 nm light and the black lines show the results tested with 850 nm light. The average optical transmissions of 100 nm and 200 nm aluminum coatings are $10^{-4}$ and $10^{-9}$, respectively. The dispersion of the optical transmission of 200 nm aluminum film is wider than that of the 100 nm aluminum film. Some pixels have high optical transmission due to the pinhole phenomenon in the aluminum film, which was also found on CCD [13].

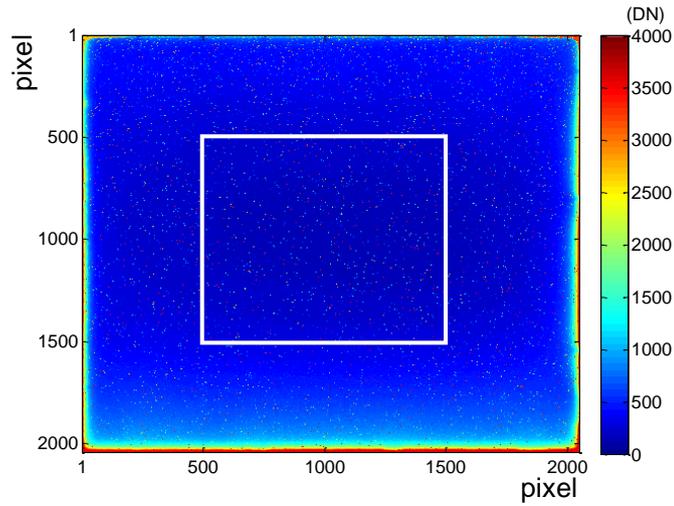

**Figure 4.** With a 660 nm light source, the image taken by the 200 nm fully Al-coated sCMOS sensor with the exposure time 1 s.

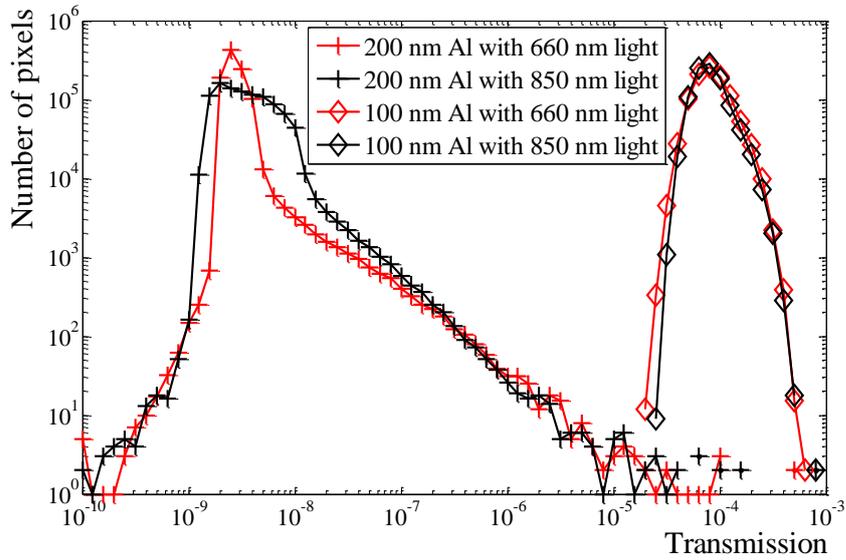

**Figure 5.** Optical transmission of 100 nm and 200 nm aluminum film tested on the central 1000 × 1000 pixels at wavelengths of 660 nm and 850 nm.

## 4. Basic performance

We investigated the basic performance of the Al-coated sCMOS sensors to verify whether there are any differences before and after the aluminum coating. These sCMOS sensors were tested in a dark environment at different temperatures and were irradiated with an $^{55}$Fe X-ray source at -15°C. The HDR mode was used in the test. Meanwhile, a standard uncoated GSENSE400BSI CMOS sensor was used as a reference.

**4.1 Readout noise and fixed pattern noise**

The readout noise is derived from 100 dark frames collected with the shortest integration time of 30 μs. The standard deviation of the signal in each pixel is calculated, and the median of the standard deviation over all pixels gives the readout noise. At the same time, the mean of the signal in each pixel is calculated and the spatial variance of these pixel-by-pixel mean values gives the fixed pattern noise (FPN). Figure 6 shows the histograms of the readout noise and the FPN of the sCMOS sensor fully coated with 200 nm Al tested at -16°C. Figure 7 shows the readout noise and the FPN of these sCMOS sensors at different temperatures. The readout noise is about 2 e- for all of the sCMOS sensors, and there is no noticeable change after coating with the aluminum films. The FPN of the sCMOS sensors is below 3 elections and the fully-aluminized sensors are a little bit better than the others.

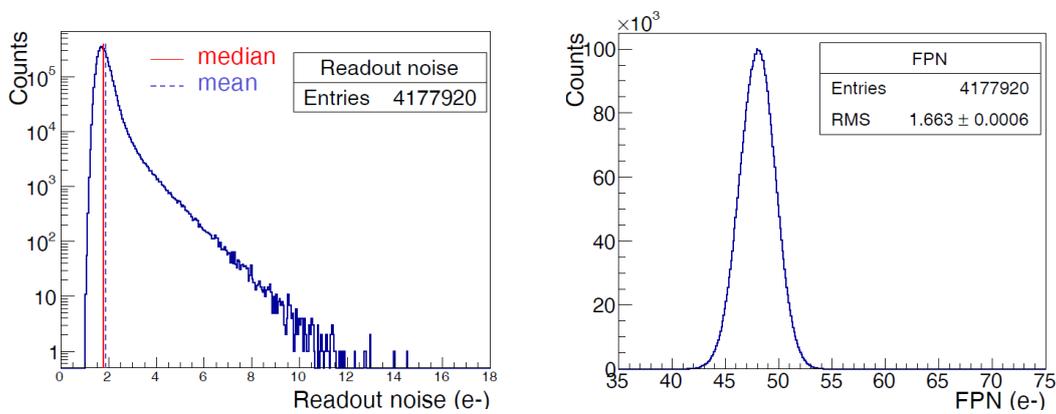

**Figure 6.** Histograms of the readout noise and the FPN of the sCMOS sensor fully coated with 200 nm Al at -16°C.

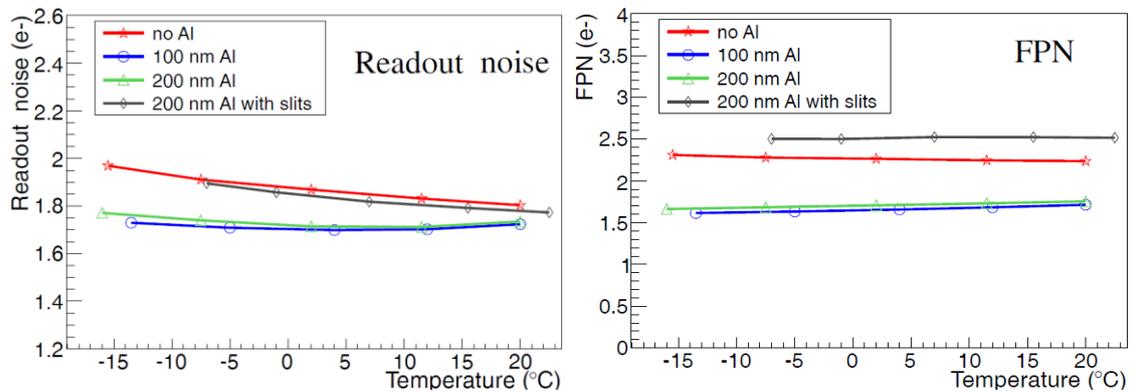

**Figure 7.** The readout noise (left) and the FPN (right) as a function of the temperature.

**4.2 Dark current**

Under the dark environment, the slope of the relationship between signal value and exposure time is calculated pixel by pixel. The median of the slopes over all pixels is used to represent the dark current of the sCMOS sensor. Figure 8 shows the dark current distribution of the sCMOS sensor fully coated with 200 nm Al at -16°C. The dark current is relatively high in some areas around the sCMOS sensor, as shown on the left of Figure 8, which is caused by the design of the circuit distribution of the sensor. Figure 9 shows the dark currents of these sCMOS sensors under different temperatures. The red line shows the dark current of the standard sCMOS sensor without the aluminum coating. The blue line shows the dark current of the sCMOS sensor fully coated with the 100 nm aluminum film. The green and black lines show the dark currents of the sCMOS sensors coated with the 200 nm aluminum film without or with slits. These results show that the dark currents of the Al-coated sensors are consistent in spite of the different thicknesses or patterns of the aluminum layers. However, at room temperature, the dark currents of the Al-coated sensors are 20 times larger than the sCMOS sensor without Al coating. Such a difference disappears when the temperature drops to about -15°C. A possible reason is that the aluminum coating changes the surface state of the sCMOS sensor, resulting in the change of the dark current.

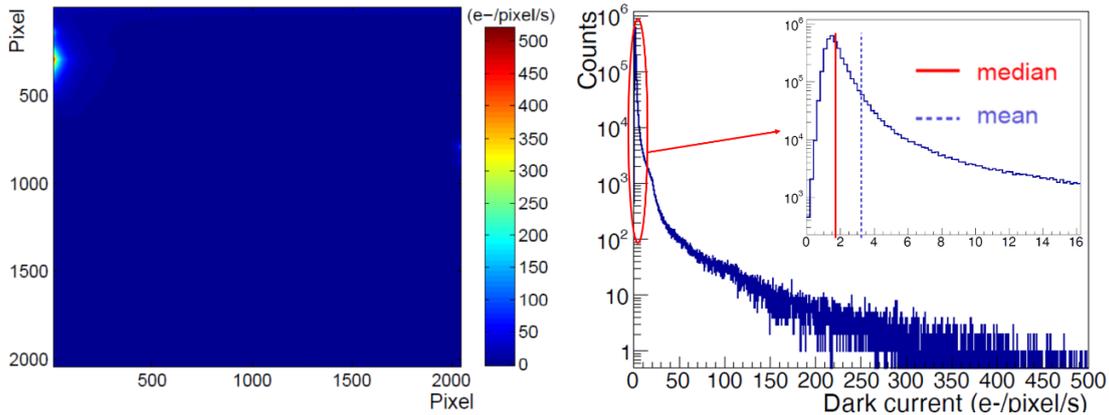

**Figure 8.** The dark current distribution of the sCMOS sensor fully coated with 200 nm Al at -16°C. Left: The image of the dark current. Right: The histogram of the dark current.

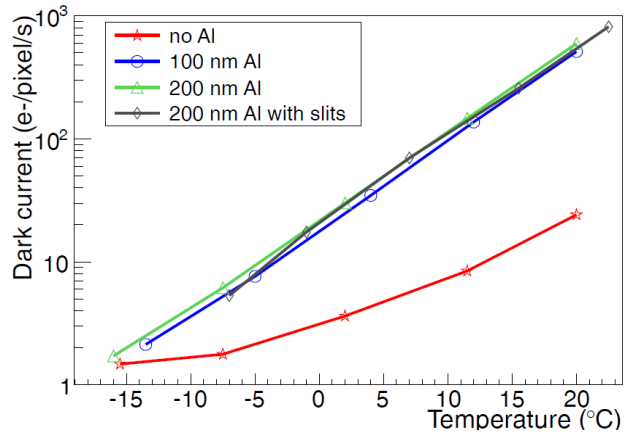

**Figure 9.** The dark current as a function of temperature.

### 4.3 X-ray spectrum

An $^{55}$Fe X-ray source was used to test the X-ray response of these sCMOS sensors. The details of the data reduction method can be found in Ref [14]. The Si escape peak, Mn $K_\alpha$ and Mn $K_\beta$ can be seen in the spectrum of the $^{55}$Fe source. Using the relationship of these positions of the peaks and their corresponding energy, a linear fitting is used to obtain the conversion gain of the sCMOS sensor. Figure 10 shows the X-ray spectra tested at -15°C. Panel (a) of Figure 10 shows the energy spectra of all X-ray events, and panel (b) shows the energy spectra of single-pixel events, of which the electrons produced by X-ray photons are collected by only one pixel. Panel (c) shows the energy spectra obtained with X-ray events collected by more than one pixel. It can be found that peak 1 in panel (a) of Figure 10 mainly comes from multi-pixel events, while peak 2 and peak 3 are mostly from single-pixel events. Because these GSENSE400BSI sensors are not fully depleted devices and some of the electrons produced in the epitaxial layer are lost during charge collection, peak 1 of the multi-pixel events shows noticeable energy loss in the spectrum. The detailed spectral analysis is beyond this paper and we will focus on the changes between the coated and uncoated sensors. The performance of these spectra is shown in Table 1. These results prove that the X-ray characteristics of the Al-coated sCMOS sensors do not change significantly at the low temperature compared with the uncoated sCMOS sensor.

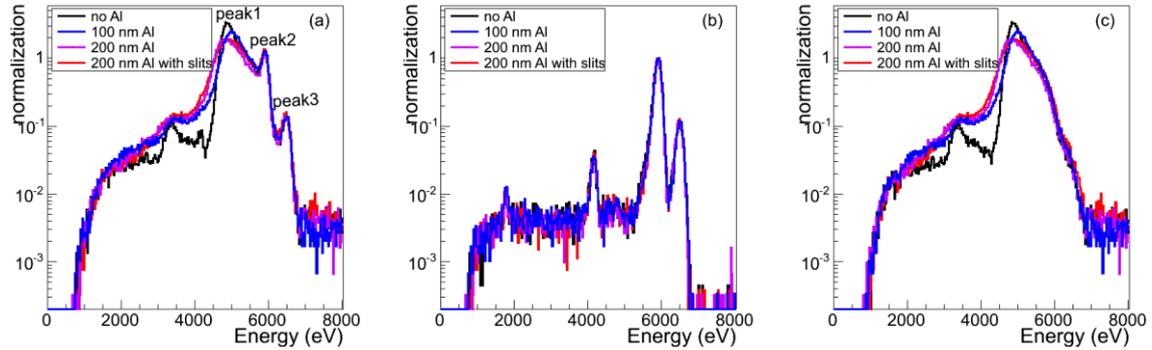

**Figure 10.** $^{55}$Fe X-ray spectra after normalization. (a) all X-ray events; (b) only single pixel events; (c) X-ray events of multiple pixels.

**Table 1.** X-ray energy spectrum characteristics of the different sCMOS sensors at -15°C.

| sCMOS sensor | Ratio of single pixel events to total X-ray events (%) | Conversion gain (eV/DN) | Energy resolution (eV) |
|---|---|---|---|
| No Al | 11.2 ± 0.2 | 2.07 ± 0.01 | 195.8 ± 1.5 |
| 100 nm Al | 10.9 ± 0.2 | 2.06 ± 0.01 | 181.6 ± 1.3 |
| 200 nm Al | 10.5 ± 0.2 | 2.28 ± 0.01 | 176.8 ± 1.4 |
| 200 nm Al with slits | 11.1 ± 0.2 | 1.94 ± 0.01 | 175.9 ± 1.4 |

### 4.4 Reliability test

To verify the reliability of the Al coatings, thermal cycle tests were conducted on the two patterns of 200 nm Al-coated sensors. For the first group, two sCMOS sensors were kept at a low temperature of -55°C for 48 hours, and then underwent 80 times thermal cycles from -55°C to 85°C. For the second group, another two sCMOS sensors were placed at a high temperature of 125°C for 48 hours, and then underwent 80 times thermal cycles from -55°C to 85°C. The dark currents during different experiment stages are shown in Figure 11. The results show that the dark current increased to less than 1.5 times.

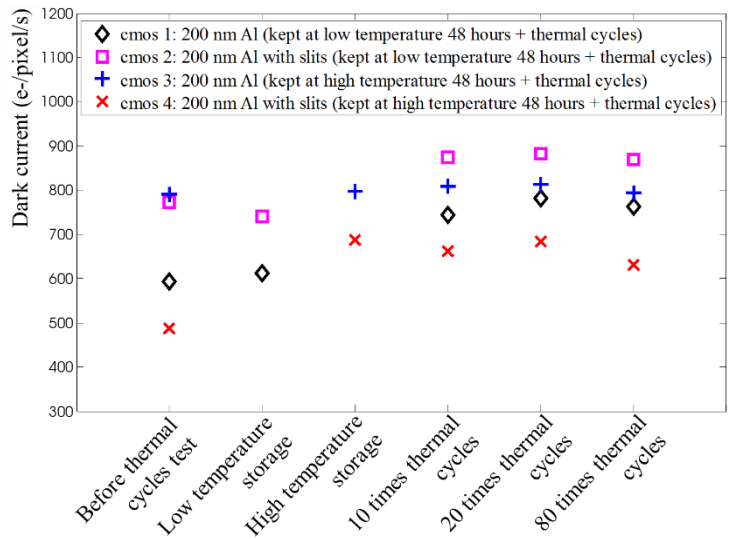

**Figure 11.** Dark currents measured at room temperature at different experiment stages.

The aging experiment of the 200 nm fully Al-coated sCMOS sensor was also carried out. The sCMOS sensor worked continuously for 15 days in a high temperature environment of 85°C. The dark currents of the sCMOS sensor before and after the aging experiment are shown in Figure 12. After the sCMOS sensor worked in the high temperature environment, the dark current of the sCMOS sensor increased to about 1.36 times. The increase of the dark current is within the range of expectation. After the thermal cycles and aging experiments, these sCMOS sensors show no visible cracks on aluminum film, demonstrating that the Al-coating could survive in the harsh environment.

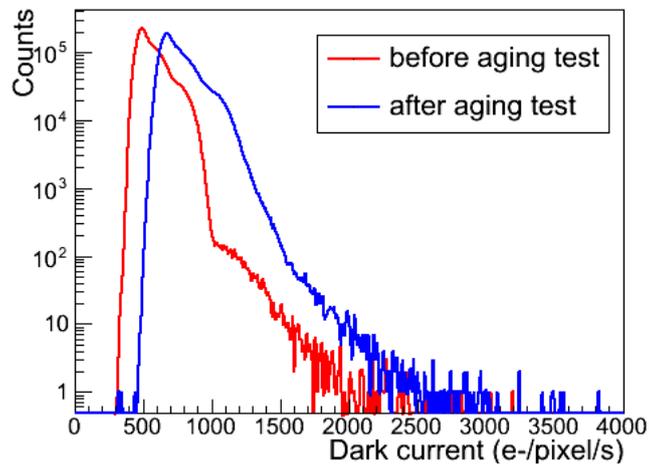

**Figure 12.** Dark current measured at room temperature before and after the aging test.

## 5. Conclusion

We designed and fabricated three types of aluminum optical blocking filters which are directly deposited on the surface of the back-illuminated sCMOS sensors. Considering the different coefficient of thermal expansion between the aluminum and semiconductor materials, two patterns of aluminum coating are used: aluminum film with full coverage and aluminum film with slits. The optical transmission measurement results show that the optical transmissions reach a level of $10^{-9}$ for 200 nm aluminum coating and of $10^{-4}$ for 100 nm aluminum coating at wavelengths of 660 nm and 850 nm. The light leakage is found around the sidewall of the sCMOS sensor even with a 70 μm aluminum film extension on each side. At room temperature, the dark currents of the Al-coated sCMOS sensors are 20 times higher than that of the sCMOS sensor without coating. However, such a difference disappears when the temperature drops to about -15°C. The readout noise, FPN, conversion gain, and energy resolution of these Al-coated sCMOS sensors are similar to that of the uncoated sCMOS sensor. After the thermal cycle and aging tests, no obvious crack on the aluminum film was found, even with the fully aluminized sCMOS sensor, and the dark current of the sCMOS sensors increased to less than 1.5 times, which is as expected. Based on these results, a full 200 nm Al coating directly deposited on a large format sCMOS sensor is designed, and the coating extension of more than 1 mm on each side is used to avoid the edge effect of light leakage.

## Acknowledgments

This work is supported by the National Natural Science Foundation of China (grant Nos. 12173055, 12273073) and the Chinese Academy of Sciences (grant Nos. XDA15310100, XDA15310300, XDA15052100).